\documentclass[12pt]{article}
\usepackage{amsmath,amsthm, amssymb}
\usepackage{graphicx}
\usepackage{leqno}
\usepackage{hyperref}
\numberwithin{equation}{section}
\begin{document}
\title{\vskip-40pt On the Analysis of Bell's 1964 Paper by Wiseman, Cavalcanti, and Rieffel}
\author{Edward J. Gillis\footnote{email: gillise@provide.net}}
				
\maketitle

\begin{abstract} 

\noindent 
In a  recent series of papers Wiseman, Cavalcanti, and Rieffel have outlined and contrasted two different  
views about what we now call Bell's theorem.  They also assert that Bell presented these two different 
versions at different times. This view is clearly at odds with the detailed explanation that Bell himself 
gave in his later writings. A careful examination of the historic 1964 paper in context shows clearly that 
Bell's own later interpretation is the correct one.

\end{abstract}

               \indent \indent \indent Einstein locality

\section{Introduction}

Just over 50 years ago Bell wrote the paper\cite{Bell_1} in which he derived the following 
result:\begin{quotation}\noindent
              "In a theory in which parameters are added to quantum mechanics to determine 
              the results of individual measurements, without changing the statistical predictions, 
              there must be a mechanism whereby the setting of one measurement device can influence 
              the reading of another instrument, however remote. Moreover, the signal involved must 
              propagate instantaneously, so that such a theory could not be Lorentz invariant."
              \end{quotation}
A half century later, there is still contentious debate about the implications of Bell's work. 

In an effort to improve communication between two 'camps' of researchers in quantum foundations, 
Wiseman, Cavalcanti, and Rieffel (WCR) have described two different ways in which Bell's result can be 
obtained\cite{Wiseman_1,Wiseman_2,Wiseman_3}\footnote{The first paper cited was authored solely by 
Wiseman, the second by Wisemann and Cavalcanti, and the third, which was a response to Norsen's 
comment\cite{Norsen_on_Wiseman}, was authored by Wiseman and Rieffel.}. They argue that one method of 
derivation is favored by a group that they call 'operationalists', and the other by a group labeled as 
'realists'. They also make the much more controversial historical claim that Bell, himself, presented 
these two different versions at different times. In their view, the earlier (1964) paper contains the 
"operationalist" version, and the "realist" version was not published until 1976\cite{Bell_1976}. Norsen 
has written a vigorous  dissent\cite{Norsen_on_Wiseman} from their interpretation of the original (1964) article. 
Although Norsen has already made a strong case, the historic importance of Bell's 1964 contribution makes it 
worth presenting some additional evidence for his point of view. 

Norsen has called into question the WCR characterization of their favored version of Bell's 1964 argument as 
'operationalist', since the formalization of some key principles (like 'parameter independence') that they 
attribute to that group  requires reference to an an ontology that is rejected by most operationalists. So 
a more neutral  terminology will be used here. I will refer to the version that they advocate simply as 'WCR'. 
The interpretation of the 1964 paper that Bell, himself, presented in his later 
writings\cite{Bell_1976,Bell_Bert,Bell_LNC}, will be labeled 'JSB' (Bell's initials). 

Let us begin by outlining the two derivations described by WCR. The background assumption for both of 
the derivations is that the quantum statistical predictions (QSP) are correct. Given this premise there are 
two ways in which to derive Bell's result. One of them is to assume: (a) that the proposed theory ($\theta$) 
implies that the setting of one of the measurement instruments does not change the total probability of the 
outcomes detected by the other instrument; (b) that the proposed theory is deterministic. This method reflects 
the WCR viewpoint. The other (JSB) derivation assumes that the proposed theory implies that all physical 
processes\footnote{What is to count as a 'physical process' is to be specified by the proposed theory.} 
propagate continuously through space within the forward light cone. Together with the background 
assumption (QSP), this \textit{implies} the determinism that is assumed in the other approach. 

WCR formalize premise (a) of their interpretation as: $P_\theta\!(B|a,b,c,\lambda) = P_\theta\!(B|b,c,\lambda),$ 
where $P$ is the function that assigns probabilities to various outcomes, $\theta$ is the theory by 
which the probabilities are calculated, $B$ represents the outcome of the second measurement, $a$ and $b$ 
represent the settings of the first and second measurement instruments, and $c$ and $\lambda$ represent 
all other (possibly hidden) variables that might be relevant. The authors label this condition as 'locality' 
(the term used by Bell in his 1964 paper), but the meaning of 'locality' is precisely what is at issue in the 
debate over the historical claim about what Bell actually proved in 1964. Shimony\cite{Shimony} proposed the 
phrase, 'parameter independence' (PI). Although that terminology is also somewhat problematic, it has been 
widely adopted, and so it will be used here. 

Bell discussed the additional assumption in the JSB approach at some length in a later work\cite{Bell_LNC}. 
In that work he called the assumption 'local causality'. He used it to derive the following 'factorizability' 
condition on probabilities (in somewhat different notation): 
$P_\theta\!(A,B|a,b,c,\lambda) = P_\theta\!(A|a,c,\lambda)P_\theta\!(B|b,c,\lambda).$  Note that this expression 
is a close parallel to the formulation of the crucial condition on expectation values that he used in the 
1964 paper. He  emphasized that the formal expression should be viewed only as \textit{consequence} of 'local 
causality' - \textit{not} as a full formulation of it.

Jarrett\cite{Jarrett} showed that Bell's derived condition is logically equivalent to the conjunction of 
parameter independence (PI) and an additional constraint that Shimony\cite{Shimony} later labeled as 'outcome 
independence' (OI): $P_\theta\!(B|A,a,b,c,\lambda) = P_\theta\!(B|a,b,c,\lambda).$ Jarrett's decomposition 
helps to distinguish various ways in which a measurement outcome might be influenced by a distant 
event, and it also makes it clear that Bell's (1990) condition is strictly stronger than PI.

Now, although the formal expression presented by Bell in 1990\cite{Bell_LNC} can be decomposed into PI and OI, 
he regarded 'local causality' as a \textit{single} principle (a "unitary property" as characterized by 
WCR\cite{Wiseman_1}). This point is important in the debate about exactly what conclusion follows from the 
argument presented in 1964. So it is desirable to use a single abbreviation to label this key assumption. Since 
Bell explicitly stated that 'local causality' does \textit{not} entail determinism, the use of this phrase 
could be confusing since he had used the term 'causality' as a synonym for 'determinism' in earlier works, including 
the 1964 article. In 1990 he gave the following informal statement of the principle:
\begin{quote}\noindent  
"The direct causes (and effects) are near by, and even the indirect causes (and effects) are no further away than 
are permitted by the velocity of light."  (Bell 1990)
\end{quote}
In what follows, this principle will be characterized as  'No Superluminal Effects' (NSE), with the understanding 
that it implies continuous propagation through space. The principle of determinism (which is \textit{not} entailed 
by NSE) will be abbreviated as 'DET", and the contradiction that Bell derived in section 4 of his paper will be 
indicated as XX. 

So the shorthand representations of the two derivations are:\newline 
WCR: $ QSP\, +\, PI\, +\, DET \; \Longrightarrow \;  XX$;\newline 
JSB: $ \;\;QSP\, +\, NSE \; \Longrightarrow \; DET$;  $ \;\;\;\;QSP\, +\, NSE\, +\, DET \; \Longrightarrow \;  XX$.

Since  $PI\, +\, DET \; \Longrightarrow \;  NSE $,  and $  NSE \; \Longrightarrow \; PI$, and 
$ QSP\, +\, NSE \; \Longrightarrow \; DET$, it is easy to see that the two sets of premises are logically 
equivalent. So there does not appear to be any serious dispute about whether Bell's result \textit{could} be 
derived in different ways. The different derivations of the result can be viewed as two different 
theorems. Serious disagreement sets in with the claim that what Bell actually presented in 1964 was 
the WCR derivation, and, hence, that both of these theorems are \textit{Bell's} theorems.  

I will argue here that the JSB version of the argument is the one that Bell actually presented in 1964. 
To understand what Bell was doing in \cite{Bell_1} consider the general structure of his argument. 
As indicated above, the JSB version has two stages:\newline
 (a) $ QSP\, +\, NSE \; \Longrightarrow \; DET$;\newline
 (b) $ QSP\, +\, NSE\, +\, DET \; \Longrightarrow \;  XX$.\newline
 The first of these is the central argument of the EPR paper\cite{EPR} as understood by Bell in 
 1964.\footnote{The EPR argument uses the  more general premise of no disturbance of one subsystem by 
 a measurement  on the other. The more general premise  is, of course, implied by the assumption of 
 no superluminal action-at-a-distance, and it appears that nearly everyone understood this. For 
 additional background, the  reader  is referred to the original article\cite{EPR}, Bohr's reply to 
 it\cite{Bohr_EPR}, Fine's Stanford Encyclopedia article\cite{Fine}, Einstein's later 
 writings\cite{Einstein_1948,Einstein_1949}, and the 1957  article by Bohm and Aharonov\cite{Bohm_Aharonov}.}  
 Bell took the \textit{validity} of this  argument  as being both well established and familiar to  his audience. 
 By conjoining the consequence  (determinism) of (a) with the two premises and deriving a contradiction, he was 
 able to show that this  version of the  EPR argument is \textit{unsound}. In other  words, either QSP or NSE 
 is false; either quantum  theory is incorrect,  or there are real physical effects that are propagated outside 
 the light cone. 
 
In fact, Bell's demonstration is a conventional proof by contradiction. (His argument is contained mainly in 
section 4, entitled "Contradiction".) Consider that in virtually every proof of this sort the author states 
the premises, demonstrates the conflict, and concludes that one of the premises must be false. Immediately 
after the passage quoted above in which he states that NSE is inconsistent with QSP, he raises the possibility 
that QSP is wrong:  
 \begin{quotation}\noindent
             ``Moreover, the signal involved must propagate instantaneously, so that such a theory 
              could not be Lorentz invariant.
              \newline \indent Of course, the situation is different if the quantum mechanical 
              predictions are of limited validity.''\end{quotation}
Bell's description of the possible violation of NSE might appear a little loose by current standards\footnote{Since 
quantum effects do not allow 'signaling', and some hypothetical superluminal effects are consistent with 
Lorentz invariance\cite{Maudlin}.}, but its meaning is clear. The perfect correlations between distant measurement 
outcomes described by EPR cannot be explained except by superluminal effects. The only possibility of avoiding 
such effects is through the failure of quantum theory.

The next section will present the case for the JSB interpretation of Bell's 1964 paper in more detail.

\section{What Bell Meant by 'Locality'}

The key point in dispute centers on what Bell meant by the term 'locality'. Bell, himself, stated later 
quite clearly that the argument presented in his 1964 paper was the JSB version. In a 1981 
essay\cite{Bell_Bert}(p. 143)\footnote{For Bell's papers that are reprinted in \textbf{Speakable and 
Unspeakable in Quantum Mechanics}, revised edition (2004), all page references here are to that edition.} 
he says: 
\begin{quote}\noindent  
"It is important to note that to the limited degree to which \textit{determinism} plays a role in the EPR 
argument, it is not assumed but \textit{inferred}. What is held sacred is the principle of 'local causality' - 
or 'no action at a distance'."\end{quote}
A few sentences later, in a footnote, he says:
\begin{quote}\noindent  
    "My own first paper on this subject[*] starts with a summary of the EPR argument 
    \textit{from locality} to deterministic hidden variables. But the commentators have almost 
    universally reported that it begins with deterministic hidden variables"\end{quote}
In \cite{Wiseman_1} (p.17) Wiseman insists that Bell's explanation of what he meant was mistaken: 
\begin{quote}\noindent  
In any case, it seems that once Bell had explicitly defined  LC[[i.e., the JSB notion of locality]], 
he wished all previous localistic notions he had used, in particular the notion of locality as per 
Definition 9[[i.e., 'locality' = PI]], to be forgotten. Moreover, after a few years 
he became convinced that it was the notion of LC that he had in mind all along."\end{quote}

Since WCR are inclined to discount the explanations that Bell offered later, let us look at a discussion 
of 'locality' given by Bell in a paper that Wiseman very aptly describes as a prequel\footnote{The paper 
on hidden variables was written prior to the EPR paper, but published later.} to "On the Einstein-Podolsky-Rosen 
paradox". In his paper on hidden variables\cite{Bell_2} Bell had shown that the requirements imposed by 
von  Neumann\cite{von_Neumann} to rule out the possibility of adding hidden variables to quantum theory were 
arbitrary and unreasonable. In the last section of the paper he proposed a requirement that he considered much 
more natural, and in which he described the \textit{motivation for considering theories of deterministic hidden 
variables}. Near the beginning of this section entitled "Locality and separability" he says: 
\begin{quote}\noindent
"... there \textit{are} features which can reasonably be desired in a hidden variable scheme. The hidden 
variables should surely have some spacial significance and should evolve in time according to prescribed laws. 
These are prejudices, but it is just this possibility of interpolating some (preferably causal) space-time 
picture, between preparation of and measurements on states, that makes the quest for hidden variables interesting 
to the unsophisticated[\textit{reference to the Einstein work cited three times in the opening 
paragraphs of the subsequent paper}]." \end{quote} 
The phrases, "spacial significance" and "interpolating some... space-time picture between", are clearly 
meant to convey the idea that "Locality" (the title of the section) includes in an essential way the notion 
that all physical processes propagate continuously through space. The term "unsophisticated" appears to be 
an ironic reference to Einstein and his resistance to the "orthodox" interpretation promoted by 
Bohr\cite{Bohr_EPR,Bohr_2} and Heisenberg\cite{Heisenberg_a,Heisenberg_b}. 

In fact, the final sentence in the quotation can be read as a very brief summary of the EPR argument. 
The possibility of maintaining a picture in which physical processes propagate continuously within 
the light cone is what leads "unsophisticated" people like Einstein (and Bell himself) to consider 
a theory of hidden variables, since without such variables one is forced to accept action-at-a-distance. 
Note the parenthetic phrase, "\textit{preferably} causal". 'Causal' is being used here, as in the subsequent, 
paper as a synonym for 'deterministic'. The qualifier, "preferably", indicates that Bell does not insist on a 
deterministic account; it is just that it is the only way to save Einstein locality and still reproduce the 
perfect correlations discussed by EPR. 

It is only after explaining why \textit{Einstein's principle of locality} (i.e., no superluminal effects) requires 
deterministic hidden variables that Bell goes on to discuss Bohm's theory\cite{Bohm}, and to characterize it as 
"grossly non-local". 

The description in this passage about the continuous propagation of physical processes through space with the 
implied consistency with relativity (given the reference to Einstein's views) is a close parallel to 
Bell's 1990 description of the principle cited above: 
\begin{quote}\noindent  
"The direct causes (and effects) are near by, and even the indirect causes (and effects) are no further away than 
are permitted by the velocity of light."  (Bell 1990)
\end{quote}

So prior to the 1964 paper Bell had already described a concept essentially equivalent to what he called 
'local causality' in 1990. Since he had also linked it to fairly well-known writings of Einstein, it was 
completely reasonable for him to assume that the concept would be familiar to his audience. Despite these 
considerations, WCR still maintain that this was \textit{not} the concept that he referred to in 
\cite{Bell_1}. Let us examine their arguments.

The principal evidence for their claim that Bell \textit{defined} 'locality' as \newline  
PI, $P_\theta\!(B|a,b,c,\lambda) = P_\theta\!(B|b,c,\lambda),$ consists of four passages from \cite{Bell_1}. 
(Some of their additional arguments will be reviewed later.)

(1) "It is the requirement of locality, or more precisely that the result of a
measurement on one system be unaffected by operations on a distant system
with which it has interacted in the past, that creates the essential difficulty." (p.14)

(2) "Now we make the hypothesis[*], and it seems one at least worth considering, that if the two measurements 
are made at places remote from one another the orientation of one magnet does not influence the result 
obtained with the other." (p.14/15)

(3) "The vital assumption[*] is that the result B for particle 2 does not depend on
the setting \textbf{a} of the magnet for particle 1, nor A on \textbf{b}." (p.15)

(4) "In a theory in which parameters are added to quantum mechanics to determine 
the results of individual measurements, without changing the statistical predictions, 
there must be a mechanism whereby the setting of one measurement device can influence 
the reading of another instrument, however remote." (p.20)

Note first that nowhere does Bell use the term, 'define' or a close equivalent, or write down any expression 
that resembles PI. (The phrase in the first passage, "or more precisely that..." is used to point out 
a particular consequence of locality - not to present a definition.) So the WCR claim is based on 
their insistence that the terms, '(un)affected', 'influence', and 'depend' \textit{must} be understood 
as implying that the \textit{total} probability of an outcome of a measurement made on one branch of an entangled 
system is altered by changes in the \textit{setting} of the measurement apparatus that acts on the other branch. 
This interpretation rests essentially on the fact that, in all of these passages, Bell refers to the dependence of 
the outcome of the second measurement \textit{exclusively} on the setting of the first instrument, rather than on 
\textit{both} the setting of the instrument and the outcome. In fact, there is a very good reason that
Bell focussed on this particular aspect of the experiment. This is explained below, but, first, it is important 
to identify the fundamental interpretive error made by WCR.

As pointed out in the Introduction, Bell's factorizability condition is logically equivalent to 
the conjunction of PI and OI. For reference these conditions are: \newline
FACT: $P_\theta\!(A,B|a,b,c,\lambda) = P_\theta\!(A|a,c,\lambda)P_\theta\!(B|b,c,\lambda).$ \newline
PI: $\;\;\;P_\theta\!(B|a,b,c,\lambda) = P_\theta\!(B|b,c,\lambda)\;\;\;\;\;\;\;\;$ 
(OI): $\;\;\;P_\theta\!(B|A,a,b,c,\lambda) = P_\theta\!(B|a,b,c,\lambda).$ 
Quantum theory clearly violates factorizability, and therefore, it violates OI. Note that the negation of  
\textit{either} PI or OI would allow a dependence of $P(B)$ on the setting, $a$. The only reason that PI appears 
to imply the independence of $P(B)$ from $a$ is that PI deals only with the \textit{total} probability. One cannot 
claim that the setting of one measurement apparatus does not "affect" or "influence" the outcome of a distant 
measurement unless both PI and OI are respected. (This is why the terminology, 'Parameter Independence', 
is problematic.) The attempt by WCR to read these very general terms as applying only to the total probability 
of the distant outcome is forced and unnatural. If taken seriously, it would simply rule out the possibility 
of \textit{explaining the correlations} between measurement outcomes, which is the whole point of the 
EPR argument.

To make this point in another way, consider that in orthodox quantum mechanics the outcome of a measurement on 
one of a pair of entangled systems is influenced by two factors that are not necessarily connected to anything 
in the past light cone of that measurement outcome: (1) the setting of the distant instrument \textit{and} 
(2) the outcome of the distant measurement. Now, "orthodox" quantum mechanics can be understood in 
either of two ways, but this statement holds true in both of them. In the "von Neumann" version the wave 
function, regarded as a genuine physical entity, undergoes a collapse that spans a spacelike interval. Viewed 
in this way, the state to which the wave function collapses (and, hence, the outcome of the second measurement) 
is \textit{nonlocally influenced} by both the setting of the measurement instrument, which reduces the number 
of possible resultant states from infinity to two, and the outcome of the first measurement, which determines 
which of the two possibilities is realized. In the operational or "Bohr-Heisenberg" version there is nothing 
physical for the measurement setting, by itself, to influence. The only meaningful influences are on experimental 
outcomes, and these are influenced (nonlocally) by both the setting of the first instrument and by the outcome 
of the first measurement. (As Norsen points out, in the purely operational version it is not even clear that one 
can formulate expressions for PI and OI.)

It remains to explain why Bell focuses solely on the setting of the apparatus. The reason is that it is the one 
aspect of the experimental arrangements envisioned that can be placed \textit{unambiguously} outside the past light 
cone of the second measurement. To see this, recall that there were two theories with which Bell was thoroughly 
familiar that could reproduce all of the statistical predictions of quantum theory. The first of these was orthodox   
quantum mechanics; the other was Bohmian theory\cite{Bohm}. As just noted, in orthodox quantum mechanics 
both the measurement setting and outcome influence the distant result. In Bohm's theory, however, it 
is \textit{only} the setting of the instrument that can be placed strictly outside the past light cone of the 
second measurement. The trajectory of the first particle (and, hence, the deterministic result of the measurement) 
is heavily influenced by the previous interaction that generated the entanglement with the second system, and this 
interaction is clearly \textit{within} the past light cone of the second measurement. So inclusion of possible 
influences of the \textit{result} of the first measurement would have involved a complicated mix of factors 
affecting the second outcome and might have detracted from the clarity of Bell's result. Thus, it was entirely 
natural, or even essential, for Bell, in searching for a clear-cut test to demonstrate the nonlocality of quantum 
theory, to frame the issue in terms of the influence exerted by the setting of the first measurement apparatus. 
This point can be driven home by considering the closing paragraph of the 1964 paper in which he describes specific 
experimental tests.
\begin{quotation}\noindent
 	"Of course, the situation is different if the quantum mechanical predictions are of limited validity. 
 	Conceivably they might apply only to experiments in which the settings of the instruments are made 
 	sufficiently in advance to allow them to reach some mutual rapport by exchange of signals with velocity 
 	less than or equal to that of light. In that connection, experiments of the type proposed by Bohm and 
 	Aharonov[*], in which the settings are changed during the flight of the particles, are crucial."
 	\end{quotation}

In further support of this explanation for his choice of phrasing, recall that Bell was hugely influenced by 
Bohm's theory\cite{Bohm}. It was Bohm's theory (apparently) that first convinced him that von Neumann's no-hidden 
variable theorem\cite{von_Neumann} was seriously flawed, and, after recognizing the nonlocal nature of the theory, 
he had spent a great deal of effort in trying to construct a version without this problematic feature. It was 
largely this effort that led him to his 1964 theorem. With this background, it is entirely understandable that 
he emphasized how the \textit{setting} of one  measurement instrument influences the outcome of the distant 
measurement, since this is the one clearly identifiable feature that is not influenced by events in the past 
light cone of the distant measurement. 

So the first three passages cited above should be read, not as a definition of 'locality', but as 
a completely unambiguous criterion for ascertaining whether the principle of locality is violated. The final 
passage is the statement that any theory consistent with QSP violates this criterion. 

In his comment Norsen\cite{Norsen_on_Wiseman} has also argued that Bell's phrasing amounts to stating 
a criterion for violating locality, rather than offering a definition. He has made a number of 
other compelling points that are  worth reviewing here. These concern Bell's brief introductory section and the 
first two paragraphs of his section 2 (Formulation). The first three passages cited above are all contained in 
this portion of the paper. As Norsen points out, in all three of these passages Bell refers to a principle enunciated 
by Einstein regarding the issues of locality and separability\cite{Einstein_1949}:
\begin{quote}
"But on one supposition we should, in my opinion, absolutely hold fast: the real factual situation of 
the system $S_2$ is independent of what is done with the system $S_1$, which is spatially separated from the former."
\end{quote}
These references are indicated by the footnotes ([*]) in the passages. The footnote for the first passage 
modifies a use of 'locality' just prior to the text quoted. In the third paper\cite{Wiseman_3} Wiseman and Rieffel 
use the term "interruption" for these references, and describe them as appeals to authority. They argue that it 
would have been better to omit them in order to "improve the grammatical and scientific clarity of the sentence". 
But they cannot rewrite Bell's paper in order to eliminate portions that conflict with their interpretation of it. 
Bell referred to the same quotation from Einstein three times in the first page and a half of his paper, directly 
qualifying the three critical passages that have been cited above. Recall from the earlier discussion that there 
were \textit{additional references to the same passage} in the prequel\cite{Bell_2} in the final section entitled 
"Locality and separability". Obviously, Bell was trying to convey to the reader the critical connotations of his 
notion of locality. 

Norsen also points out that Wiseman's interpretation of 'locality' as PI is inconsistent with Bell's statement 
at the beginning of the paper that additional (i.e.., hidden) variables were needed to \textit{restore}  
locality to quantum theory. (Maudlin has also made this point as reported by Wiseman in \cite{Wiseman_1}). PI 
is a central feature of quantum mechanics; it is what prevents superluminal signaling within the theory. It is 
very difficult to believe that Bell was unaware of such a basic property of quantum theory, or that he thought 
that Einstein was unaware of it.

To a large extent, the dispute about what Bell proved in 1964 centers on the adequacy of his recapitulation of 
the EPR argument in the opening paragraph of his second section, "Formulation". Because it is crucial to the 
issue at hand it is worth quoting in full. 
\begin{quotation}\noindent
               ``With the example advocated by Bohm and Aharonov[*], the EPR argument is the following.
                 Consider a pair of spin one-half particles formed somehow in the singlet state and moving 
                 freely in opposite directions. Measurements can be made, say by Stern-Gerlach magnets, on
                 selected components of the spins $\sigma_1$ and $\sigma_2$. If measurement of the component 
                 $\sigma_1\cdot a$, where a is some unit vector, yields the value +1 then, according to quantum 
                 mechanics, measurement of $\sigma_2\cdot a$ must yield the value -1 and vice versa. Now we make 
                 the hypothesis[[reference to Einstein quotation]], and it seems one at least worth considering, 
                 that if the two measurements are made at places remote from one another the orientation of one 
                 magnet does not influence the result obtained with the other. Since we can predict in advance 
                 the result of measuring any chosen component of $\sigma_2$, by previously measuring the same 
                 component of $\sigma_1$, it follows that the result of any such measurement must actually be 
                 predetermined. Since the initial quantum mechanical wave function does not determine the result 
                 of an individual measurement, this predetermination implies the possibility of a more complete 
                 specification of the state."
              	\end{quotation}
              	
The general form of the argument has already been briefly summarized above. Given the statistical predictions of 
quantum theory, there are systems with physical quantities that are not determined by the theory, but which can 
be precisely ascertained by making measurements on entangled partner systems that are spacelike separated. If we 
assume that there is no action-at-a-distance then these quantities must be determined by events in the common past 
of both systems. Since quantum theory does not provide specific values for all of these quantities it is incomplete.
(A complete theory would yield values for all such quantities, and would, therefore, be deterministic.)

For the reasons given above WCR insist that Bell is not using Einstein's notion of locality here (no 
action-at-a-distance), but rather the much weaker assumption of parameter independence. Therefore, they 
conclude that Bell is mistaken in assuming that determinism follows from the stated premises. In other words, 
despite the fact that he states that he is summarizing the EPR argument, Bell, without fully realizing it, 
substitutes a weaker premise for Einstein's concept of locality, and then, without fully realizing it, makes 
an assumption of determinism rather than simply restating an established result. 

Without a truly compelling reason for accepting the WCR interpretation of 'locality', their reading of this 
paragraph appears exceptionally strained and artificial. We have already seen that the argument for that  
interpretation evaporates on closer examination, and also that Bell had previously used the term 'locality' 
as essentially synonymous with Einstein's principle of no action-at-a-distance. But, since Norsen has also stated 
that this very brief discussion could benefit from a more general description of 'locality', it will be helpful to 
view Bell's short summary of EPR against the most relevant background. 

As Bell states, he was working from the version of the EPR "paradox" that had been presented by Bohm and 
Aharonov\cite{Bohm_Aharonov} in 1957 (only seven years before his paper). The closing paragraphs of their 
introductory section summarize the EPR argument and apply it to their proposed experimental arrangement. 
This straightforward exposition should make it clear why Bell did not feel compelled to repeat it at greater length.
\begin{quotation}\noindent
"One could perhaps suppose that there is some hidden interaction between $B$ and $A$, or between $B$ and 
the measuring apparatus, [[which measures $A$]] which explains the above behavior. Such an interaction 
would, at the very least, be outside the scope of current quantum theory. Moreover, it would have to be 
instantaneous, because the orientation of the measuring apparatus could very quickly be changed, and the 
spin of $B$ would have to respond immediately to the change. Such an immediate interaction between distant 
systems would not in general be consistent with the theory of relativity. 
\newline \indent This result constitutes the essence of the paradox of Einstein, Rosen, and Podolsky."\end{quotation}

This passage clearly describes the apparent conflict between the predictions of quantum theory and the 
principle of no action-at-a-distance. It also makes obvious the concern that violations of this principle 
would occur if the \textit{setting} of one of the measurement instruments (which could be changed while the 
entangled particles were in flight) affected the distant measurement. 
 
We can now turn to the other arguments offered by WCR (summarized on page 10 of \cite{Wiseman_1} and 
repeated in \cite{Wiseman_3}). In reference to the interpretation just offered Wiseman says:
\begin{quotation}\noindent 
 "To me, the advantages of this reading are demonstrably outweighed by its many
disadvantages: i) it does not explain why Bell would, in 1964, state his result \textit{four 
times} as requiring two assumptions, locality and determinism, and not once as requiring 
only the assumption of locality; ii) it does not explain why in his first subsequent paper
on the topic of hidden variables [*], after seven years to think about how best to
explain his result, he still states it (somewhat redundantly) as being "that no local
deterministic hidden-variable theory can reproduce all the experimental predictions of
quantum mechanics" [*]; iii) it does not explain why Bell would, in 1964, define locality
\textit{four times} in terms of independence from the remote setting, as per Definition 9, 
and never any other way; iv) it does not explain why Bell would state the conclusion of
the supposedly crucial first part of his theorem as being merely that it "implies the
possibility of a more complete specification of the state."; v) it does not explain why
Bell would place this supposedly crucial first part \textit{prior} to the mathematical 
formulation of his result, and not mention it anywhere else in the paper." 
 \end{quotation}  
  
The third point has been dealt with extensively above. Let us consider point (i).  
The four passages that Wiseman is referring to are:    
  \begin{quote}\noindent 
 "In this note that idea [[causality and locality]] will be formulated mathematically, and shown to be 
 incompatible with the statistical predictions of quantum mechanics." (p.14)
   \end{quote}
  \begin{quote}\noindent 
 "This is characteristic, according to the result to be proved here, of any such theory [[like Bohm's hidden 
  variable theory]] 
 which reproduces exactly the quantum mechanical predictions." (p.14)
   \end{quote}
 \begin{quote}\noindent
 "the statistical predictions of quantum mechanics are incompatible with separable predetermination."(p.20)
 \end{quote}
 \begin{quote}\noindent
              ``In a theory in which parameters are added to quantum mechanics to determine 
              the results of individual measurements, without changing the statistical predictions, 
              there must be a mechanism whereby the setting of one measurement device can influence 
              the reading of another instrument, however remote. '' (p.20)  \end{quote}  
 
 The first of these passages is the third sentence of  "On the Einstein-Podolsky-Rosen paradox". The two 
 opening sentences that precede it are:  
 \begin{quotation}\noindent
 "The paradox of Einstein, Podolsky, and Rosen[*] was advanced as an argument that quantum mechanics could not be 
 a complete theory but should be supplemented by additional variables. These additional variables were to restore 
 to the theory causality and locality[reference to Einstein's 1949 passage]." 
  \end{quotation}  
 Again, Bell is taking the EPR argument as an  established result. That argument proceeded from the no-disturbance 
 assumption implied by Einstein locality (NSE), and the limited set of quantum predictions that involve perfect 
 correlations between outcomes of spacelike-separated  entangled systems. It \textit{concluded} that 
 a hidden-variable (deterministic) theory was required. Locality  (i.e., Einstein locality) and causality (i.e., 
 determinism) are \textit{premises} of Bell's argument; WCR fail to distinguish between premises and 
 assumptions. The first of these premises, Einstein locality, is an assumption, but the  second, determinism, is 
 a property that follows from (Einstein) locality and the other assumptions of the EPR argument. This point can be 
 equally well applied to the other three passages mentioned.  
 
 The same answer can be directed to point (ii). But, in addition, the paper cited\cite{Bell_1971} includes 
 statements that clearly run counter to the interpretation that WCR try to construct. In again briefly summarizing 
 the EPR argument, Bell (p. 31) talks about filling out quantum  theory in a way that would be "manifestly local". 
 But in the view advocated by WCR, quantum theory is already  "manifestly local" since PI is one of the basic 
 properties of standard quantum theory. 
 
 The fourth point, referring to Bell's brief recapitulation of the EPR argument in the opening paragraph of Section 2,
 is based on the failure to see that  the conclusion of Bell's EPR description consists of two  sentences - not just 
 one. In the penultimate sentence Bell states very clearly that the need for a deterministic theory 
 \textit{"follows"} from the assumptions of no action-at-a-distance and the perfect correlations predicted 
 by quantum theory in the type of experiment that he discusses. In the final sentence he notes the EPR conclusion that 
 such a theory would involve a more complete specification of the state. The inference to determinism is an 
 essential premise of Bell's argument. Bell closes this recapitulation by describing the hope of 'completing' the 
 theory in a manner that allows one to save Einstein locality because this is what his subsequent demonstration 
 will show to be impossible. 
 
 Point v) reflects the insistence by WCR that in order to use the inference of the EPR argument from (Einstein) 
 locality to determinism, Bell would have needed to present an explicit logical formalization  of the 
 the argument, rather than simply assume that his audience was familiar with it. This insistence simply ignores 
 the context in which Bell was writing. The EPR  argument had come to be widely known as the EPR \textit{"paradox"}. 
 The  term, 'paradox', was used both by Bell  and by Bohm-Aharonov in the titles of their papers. The argument was 
 seen as a paradox because of the obvious  clash between the principle of no superluminal action-at-a-distance 
 (which was regarded as essential to relativity), and the perfect correlations between spacelike-separated measurement 
 results that quantum theory predicted, but  could not explain. Given this very widespread understanding, it was 
 entirely reasonable for Bell to proceed based on a brief, informal recapitulation of EPR.

\section{Summary}

Wiseman, Cavalcanti, and Rieffel deal with a number of issues in the debate between the groups that they 
characterize as 'operationalists' and 'realists', and, in particular, the various meanings that can be attached 
to the term 'locality'. These issues are well worth exploring, but the only question that is being addressed in 
this comment concerns what Bell demonstrated in his 1964 paper. 

The JSB interpretation of Bell's argument was outlined in the introduction:\newline 
(a) (EPR argument)$\;\;\;\;\;\;\;\;\;\;\;\;\;\;\;\;\;\;\;\;\;\;\;\;\;\;\; QSP\, +\, NSE \;\Longrightarrow\; DET$;\newline
(b) (Bell's argument based on EPR)  $\; \; \; \; \;  QSP\, +\, NSE\, +\, DET \; \Longrightarrow \;  XX$.\newline
The WCR version of Bell's argument was represented as:\newline 
(b') $ QSP\, +\, PI\, +\, DET \; \Longrightarrow \;  XX$.

Since (a), the EPR argument, was regarded as an established result, Bell did not attempt to formalize it. What 
was "formulated mathematically" was a \textit{consequence of the conjunction} of the premises of (b) or (b'). 
Since the conjunctions of the premises of (b) and (b') are (essentially) logically equivalent, this does not 
tell us whether Bell's use of the term 'locality' should be formalized as parameter independence or as no 
(superluminal) action-at-a-distance. The fact that determinism (DET) is used as a \textit{premise} in both versions 
fully explains why Bell refers to it in stating his conclusion. So the reference to it cannot be used in support 
of the WCR interpretation. In fact, as Norsen has pointed out, the fact that the paper is entitled 
"On the Einstein-Podolsky-Rosen paradox" argues strongly that Bell was taking determinism as a consequence 
of the EPR argument - not as an independent assumption.

So the case for interpreting 'locality' as parameter independence turns entirely on an unreasonably restrictive 
interpretation of terms like "affect" and "influence", and on the fact that Bell limited his discussion of 
possible nonlocal effects to potential influences by the \textit{setting of a measurement instrument }on 
a distant measurement outcome. The insistence that nonlocal effects must take into account both the setting of 
an instrument and the result  of the measurement made with that instrument appears to result from a failure to 
see that standard quantum theory was not the \textit{only} theory considered by Bell that was capable of 
reproducing the quantum statistical predictions. In Bohm's theory, which Bell had studied in great depth, the 
result of a measurement depends on a combination of factors, some of which are inside the past light cone of the 
distant measurement, and some of which are outside. The only event that can be placed clearly outside the past light 
cone is the setting of the instrument.\footnote{The assumption that the setting \textit{can} be securely placed 
outside the past light cone of the second measurement depends on the denial of \textit{superdeterminism}. This is 
a point that Bell, later\cite{Bell_LNC}, readily acknowledged.}

So there is no real basis for the WCR interpretation of 'locality'. In contrast, the case for reading 'locality' 
as no action-at-distance is very strong.  Bell's opening remarks that an extension of quantum theory would 
"\textit{restore} to the theory ...locality", make perfect sense, as does his recapitulation of the EPR 
argument (particularly against the background of the Bohm-Aharonov discussion). One does not need to argue away 
the three references to Einstein's 1949 passage that qualify his discussions of locality. His prior use of the 
term, 'locality', (which was consistent with the contemporary understanding) carried the clear connotation of 
continuous, subluminal propagation through space. Finally, we should consider the manner in which Bell concluded 
his analysis of the EPR argument - by emphasizing the violation of \textit{Einstein locality}.

\section*{Acknowledgements} 
I would like to thank Travis Norsen, Howard Wiseman, and an anonymous referee for their comments on this paper.

\end{document}